\documentclass[12 pt,twoside,prd,amsmath,amssymb,tightenlines,showpacs,
eqsecnum]{revtex4}
\usepackage{epsfig}
\usepackage{mathptmx}
\usepackage{hyperref}
\usepackage{bm}
\usepackage{graphicx}
\newcommand{\hnl}{\htmladdnormallink}

\renewcommand{\cal}{\mathcal}

\newcommand {\ve}{\varepsilon}

\newcommand {\cG}{\cal G}
\newcommand {\cD}{\cal D}
\newcommand {\cL}{\cal L}

\newcommand {\bp}{\bar \psi}

\newcommand {\vf}{\varphi}

\def \myfigures #1#2#3#4#5#6#7#8
{\begin{figure}[ht]
    \begin{center}
        \includegraphics[width=#2 \textwidth]{#1.eps}
        \hfill
        \includegraphics[width=#6 \textwidth]{#5.eps}
        \parbox[t]{#4\textwidth}{\caption {#3}\label{#1}}
        \hfill
        \parbox[t]{#8\textwidth}{\caption {#7}\label{#5}}
    \end{center}
\end{figure} }

\def \myfigs #1#2#3#4#5#6#7#8
{\begin{figure}[ht]
    \begin{center}
        \includegraphics[width=#2 \textwidth, angle=-90]{#1.eps}
        \hfill
        \includegraphics[width=#6 \textwidth, angle=-90]{#5.eps}
        \vskip 5 mm
        \parbox[t]{#4\textwidth}{\caption {#3}\label{#1}}
        \hfill
        \parbox[t]{#8\textwidth}{\caption {#7}\label{#5}}
    \end{center}
\end{figure}}

\def\myfigure #1#2#3#4
{\begin{figure}[ht]\begin{center}
\includegraphics[width=#2 \textwidth]{#1.eps}
\parbox[t]{#4\textwidth}{\caption{#3}\label{#1}}
\end{center}\end{figure}}

\begin{document}
\title{Spinor field and accelerated regimes in cosmology}
\author{Bijan Saha}
\affiliation{Laboratory of Information Technologies\\ Joint
Institute for Nuclear Research, Dubna\\ 141980 Dubna, Moscow
region, Russia} \email{saha@thsun1.jinr.ru}
\homepage{http://thsun1.jinr.ru/~saha}
\date{\today}
\begin{abstract}
A self-consistent system of interaction nonlinear spinor and
scalar fields within the scope of a BI cosmological model filled
with perfect fluid is considered. The role of spinor field in the
evolution of the Universe is studied. It is  shown that the spinor
field nonlinearity can generate a negative effective pressure,
which can be seen as an alternative source for late time
acceleration of the Universe.
\end{abstract}

\pacs{04.20.Ha, 03.65.Pm, 04.20.Jb, 98.80.Cq}

\maketitle

\bigskip
\section{Introduction}

The accelerated mode of expansion of the present day Universe
encourages many researchers to introduce different kind of sources
that is able to explain this. Among them most popular is the dark
energy given by a $\Lambda$ term \cite{PRpadma,sahni,lambda},
quintessence \cite{caldwell,starobinsky,zlatev,pfdenr}, Chaplygin
gas \cite{kamen,pfden}. Recently cosmological models with spinor
field have been extensively studied by a number of authors in a
series of papers \cite{sgrg,sjmp,smpla,sprd,bited,green}. The
principal motive of the papers \cite{sgrg,sjmp,smpla,sprd,bited}
was to find out the regular solutions of the corresponding field
equations. In some special cases, namely with a cosmological
constant ($\Lambda$ term) that plays the role of an additional
gravitation field, we indeed find singularity-free solutions. It
was also found that the introduction of nonlinear spinor field
results in a rapid growth of the Universe. This allows us to
consider the spinor field as a possible candidate to explain the
accelerated mode of expansion. Note that similar attempt is made
in a recent paper by Kremer {\it et. al.} \cite{kremer1}. In this
paper we study the role of a spinor field in the late-time
acceleration of the Universe. To avoid lengthy calculations, we
mainly confine ourselves to the study of master equation
describing the evolution of BI Universe. We here give the
solutions to the spinor and scalar field equations symbolically,
for details one can consult \cite{sprd,bited}.

\section{Basic equations: a brief journey}

We consider a self consistent system of nonlinear spinor and
scalar fields within the scope of a Bianchi type-I gravitational
field filled with a perfect fluid. The spinor and the scalar field
is given by the Lagrangian
\begin{equation}
\cL = \frac{i}{2} \biggl[\bp \gamma^{\mu} \nabla_{\mu} \psi-
\nabla_{\mu} \bar \psi \gamma^{\mu} \psi \biggr] - m\bp \psi + F +
\frac{1}{2} (1 + \lambda_1 F_1)\,\vf_{,\alpha}\vf^{,\alpha},
\label{lag}
\end{equation}
where $\lambda$ is the coupling constant and $F$ and $F_1$ are
some arbitrary functions of invariants generated from the real
bilinear forms of a spinor field. Here we assume $F = F (I,J)$ and
$F_1 = F_1 (I,J)$ with $I = S^2$,\, $ S = \bp \psi$,\, $J = P^2$
and $P = i \bar \psi \gamma^5 \psi$.

The gravitational field is chosen in the form
\begin{equation}
ds^2 = dt^2 - a_1^2 dx_1^2 - a_2^2 dx_2^2 - a_3^2 dx_3^2,
\label{BI1}
\end{equation}
where $a_i$ are the functions of $t$ only and the speed of light
is taken to be unity. We also define
\begin{equation}
\tau = a_1 a_2 a_3. \label{taudef}
\end{equation}

We consider the spinor and scalar field to be space independent.
In that case for the spinor, scalar and metric functions we find
the following expressions \cite{bited}.

For $F=F(I)$ we find $S=C_0/\tau$ with $C_0$ being an integration
constant. The components of the spinor field in this case read
\begin{equation}
\psi_{1,2}(t) = (C_{1,2}/\sqrt{\tau}) e^{-i\beta}, \quad
\psi_{3,4}(t) = (C_{3,4}/\sqrt{\tau}) e^{i\beta}, \label{psiI}
\end{equation}
with the integration constants obeying $C_0$ as $C_0 = C_{1}^{2} +
C_{2}^{2} - C_{3}^{2} - C_{4}^{2}.$ Here $\beta = \int(m -
{\cD})dt $.

For $F=F(J)$ in case of massless spinor field we find $P=D_0/\tau$. The
corresponding components of the spinor field in this case read:
with
\begin{eqnarray}
\psi_{1,2} &=& \bigl(D_{1,2} e^{i \sigma} + iD_{3,4}
e^{-i\sigma}\bigr)/\sqrt{\tau},\nonumber \\ \label{psiJ}\\ \psi_{3,4} &=&
\bigl(iD_{1,2} e^{i \sigma} + D_{3,4} e^{-i
\sigma}\bigr)/\sqrt{\tau},\nonumber
\end{eqnarray}
with $D_0=2\,(D_{1}^{2} + D_{2}^{2} - D_{3}^{2} -D_{4}^{2}).$

For the scalar field we find
\begin{equation}
\vf = C \int \frac{dt}{\tau (1 + 2\lambda_1 F_1)}, \quad C = {\rm
const.} \label{sfsol}
\end{equation}

Solving the Einstein equation for the metric functions we find
\begin{equation}
a_i(t) = D_{i} [\tau(t)]^{1/3} \exp \bigl[X_i \int\limits_{0}^{t}
[\tau (t')]^{-1}dt' \bigr], \label{a_i}
\end{equation}
with the integration constants obeying $$D_1 D_2 D_3 = 1, \qquad
X_1 + X_2 + X_3 = 0,$$

As one sees, the spinor, scalar and metric functions are in some
functional dependence of $\tau$. It should be noted that besides
these, other physical quantities such as spin-current, charge etc.
and invariant of space-time are too expressed via $\tau$
\cite{sprd,bited}. It should be noted that at any space-time
points where $\tau = 0$ the spinor, scalar and gravitational
fields become infinity, hence the space-time becomes singular at
this point \cite{bited}. So it is very important to study the
equation for $\tau$ (which can be viewed as master equation) in
details, exactly what we shall do in the section to follow. In
doing so we analyze the role of spinor field in the character of
evolution.

\section{Evolution of BI universe and role of spinor field}

The equation for $\tau$ is found from the Einstein one:
\begin{equation}
R_\mu^\nu - \frac{1}{2} \delta_\mu^\nu R = \kappa T_\mu^\nu +
\delta_\mu^\nu \Lambda. \label{ee}
\end{equation}
The details can be found in \cite{sprd}. This equation indeed
describes the evolution of the universe and has the following
general form:
\begin{equation}
\frac{\ddot \tau}{\tau}= \frac{3}{2}\kappa
\Bigl(T_{1}^{1}+T_{0}^{0}\Bigr) + 3 \Lambda, \label{dtau}
\end{equation}
where $\Lambda$ is the cosmological constant, $T_\mu^\nu$ is the
energy-momentum tensor. Note also that here a positive $\Lambda$
corresponds to the universal repulsive force, while a negative one
gives an additional gravitational force. Note that a positive
$\Lambda$ is often considered to be a form of dark energy. Though
our main object is to verify the role of spinor field in the
evolution of the Universe, we include the $\Lambda$ term in order
to explain some results obtained later. For this purpose we recall
that the Bianchi identity $G_{\mu;\nu}^{\nu}= 0$ gives
\begin{equation}
{\dot T}_{0}^{0} = - \frac{\dot \tau}{\tau}\bigl(T_{0}^{0} -
T_{1}^{1}\bigr). \label{conservds}
\end{equation}
After a little manipulation from \eqref{dtau} and
\eqref{conservds} one finds the following expression for $T_0^0$:

\begin{equation}
\kappa T_0^0 = 3 H^2 - \Lambda - C_{00}/\tau^2, \label{kt00}
\end{equation}
where the definition of the generalized Hubble constant $H$
as
\begin{equation}
H = \frac{1}{3}{\dot \tau}/\tau, \label{HC}
\end{equation}
Let us now stop here for a while. Consider the case when $\Lambda
= 0$. At the moment when expansion rate is zero (it might be at a
time prior to the "Big Bang", or sometimes in the far future when
the universe cease to expand we have $H = 0$. Then the
nonnegativity of $T_0^0$ suggests that $C_{00} \le 0$. Let us now
consider another case when $\tau$ is large enough for the term
$1/\tau^2$ to be omitted. As we know $T_0^0$ (the energy density),
decreases with the increase of $\tau$. If $\tau$ is big enough for
$T_0^0$ to be neglected, from \eqref{kt00} we find $$ 3 H^2 -
\Lambda \to 0.$$ It means for $\tau$ to be infinitely large,
$\Lambda \ge 0.$ In case of $\Lambda = 0$ we find that beginning
from some value of $\tau$ the rate of expansion of the Universe
becomes trivial, that is the universe does not expand with time.
Whereas, for $\Lambda > 0$ the expansion process continues
forever.  As far as negative $\Lambda$ is concerned, its presence
imposes some restriction on the energy density $T_0^0$, namely,
$T_0^0$ can never be small enough to be ignored. It imposes some
restrictions on $\tau$, precisely, there exists some
upper limit for $\tau$ (note that $\tau$ is essentially
nonnegative, i.e. bound from below). Thus we see that a negative
$\Lambda$, depending on the choice of parameters can give rise to
an oscillatory mode of expansion. Thus we come to the following
conclusion:

{\it Let $T_\mu^\nu$ be the source of the Einstein field equation;
$T_0^0$ is the energy density and $T_1^1,\,T_2^2,\,T_3^3$ are the
principal pressure and $T_1^1 = T_2^2 = T_3^3$. An ever-expanding
BI Universe may be obtained if and only if the  $\Lambda$ term is
positive (describes a repulsive force and can be viewed as a form
of dark energy) and is introduced into the system as in
\eqref{ee}.}

It should be noted that the other types of dark energy such as
quintessence, Chaplygin gas enters into the system as a part of
$T_\mu^\nu$ and corresponding energy density decreases with the
increase of the Universe, hence cannot be considered as source for
ever-expanding Universe.

Let us now go back to the Eq. \eqref{dtau}. The components of the
energy-momentum tensor read:
\begin{eqnarray}
 T_{0}^{0} &=& mS - F + \frac{1}{2} ( 1 + 2 \lambda_1
F_1) {\dot \vf}^2 + \ve_{pf}, \nonumber\\ \label{total}
\\
T_{1}^{1} &=& T_{2}^{2} = T_{3}^{3} = {\cD} S + {\cG} P - F -
\frac{1}{2} ( 1 + 2 \lambda_1 F_1) {\dot \vf}^2  - p_{pf},\nonumber
\end{eqnarray}
where, ${\cD} = 2 S dF/dI + \lambda_1 S {\dot \vf}^2 dF_1/dI$ and
${\cG} = 2 P dF/dJ + \lambda_1 P {\dot \vf}^2 dF_1/dJ.$ In
\eqref{total} $\ve_{pf}$ and $p_{pf}$ are the energy density and
pressure of the perfect fluid, respectively and related by the
equation of state  $p_{pf} = \zeta \ve_{pf}$, where $\zeta \in
[0,\,1]$.

Let us now study the equation for $\tau$ in details and clarify the
role of material field in the evolution of the Universe. For
simplicity we consider the case when both $F$ and $F_1$ are the
functions of $I\, (S)$ only. For simplicity we set $C = 1$ and
$C_0 = 1$. Note that from the Bianchi identity for $\ve_{pf}$ and $p_{pf}$
we find $\ve_{pf} = \ve_0/\tau^{1+\zeta}$ and $p_{pf} = \zeta_0
\ve_0/\tau^{1+\zeta}$. Further we set $\ve_0 = 1$. Assuming that
$F = \lambda S^q$ and $F_1 = S^r$, for the effective energy
density and effective pressure we find
\begin{eqnarray}
 T_{0}^{0} &=& \frac{m}{\tau} - \frac{\lambda}{\tau^q} +
 \frac{\tau^{r-2}}{2(2 \lambda_1 + \tau^r)} +
 \frac{1}{\tau^{1 + \zeta}} + \Lambda \equiv \ve \, \nonumber\\ \label{total0}
\\
T_{1}^{1} &=& \frac{(q -1)\lambda}{\tau^q} - \frac{[(2 - r) \lambda_1 +
\tau^r ] \tau^{r-2}}{2(2 \lambda_1 + \tau^r)^2} - \frac{\zeta}{\tau^{1+\zeta}} -
\Lambda \equiv p.\nonumber
\end{eqnarray}
Taking into account that $T_0^0$ and $T_1^1$ are the functions of
$\tau$,  only, the Eq. \eqref{dtau} can now be presented as
\begin{equation}
\ddot \tau = {\cal F}(q_1,\tau), \label{newtd}
\end{equation}
where we define
\begin{equation}
{\cal F}(q_1,\tau) =  (3/2) \kappa \Bigl(m + \lambda (q -2)\tau^{1
- q} + \lambda_1 r \tau^{r-1}/2(2 \lambda_1 + \tau^r)^2 +
(1-\zeta)/\tau^\zeta \Bigr) + 3 \Lambda \tau, \label{force}
\end{equation}
where $q_1 = \{\kappa,m,\lambda,\lambda_1,q,r,\zeta\}$ is the set
of problem parameters. The En. \eqref{newtd} allows the
following first integral:
\begin{equation}
\dot \tau = \sqrt{2[E - {\cal U}(q_1,\tau)]} \label{1stint}
\end{equation}
where we denote
\begin{equation}
 {\cal U}(q_1,\tau) = - \frac{3}{2}\Bigl[\kappa\Bigl(m \tau -
 \lambda /\tau^{q - 2} - \lambda_1 /2(2 \lambda_1 + \tau^r) +
 \tau^{1-\zeta}\Bigr) - \Lambda \tau^2\Bigr]. \label{poten}
\end{equation}
From a mechanical point of view Eq. \eqref{newtd} can be
interpreted as an equation of motion of a single particle with
unit mass under the force $\mathcal F(q_1,\tau)$. In
\eqref{1stint} $E$ is the integration constant which can be
treated as energy level, and ${\cal U}(q_1,\tau)$ is the potential
of the force $\mathcal F(q_1, \tau)$. We solve the Eq.
\eqref{newtd} numerically using Runge-Kutta method. The initial
value of $\tau$ is taken to be a reasonably small one, while the
corresponding first derivative $\dot \tau$ is evaluated from
\eqref{1stint} for a given $E$. As one sees, the positivity of the
radical imposes some restriction on the value of $\tau$, namely in
case of $\lambda > 0$ and $q \ge 2$ the value of $\tau$ cannot be
too close to zero at any space-time point. It is clearly seen from
the graphical view of the potential [cf. Fig. \ref{potenf}]. Thus
we can conclude that for some special choice of problem parameters
the introduction of nonlinear spinor field given by a self-action
provides singularity-free solutions. For numerical solutions we
set $\kappa = 1$, spinor mass $m = 1$, the power of nonlinearity
we choose as  $q = 4$, $ r = 4$ and for perfect fluid we set
$\zeta = 1/3$ that corresponds to a radiation. Here, in the
figures we use the following notations:\\ 1 corresponds to the
case with self-action and interaction, i.e., $\lambda = 1$,
$\lambda_1 = 1$;\\ 2 corresponds to the case with self-action
only, i.e., $\lambda = 1$, $\lambda_1 = 0$;\\3 corresponds to the
case with interaction only, i.e., $\lambda = 0$, $\lambda_1 = 1$.

\vskip 1 cm

\myfigs{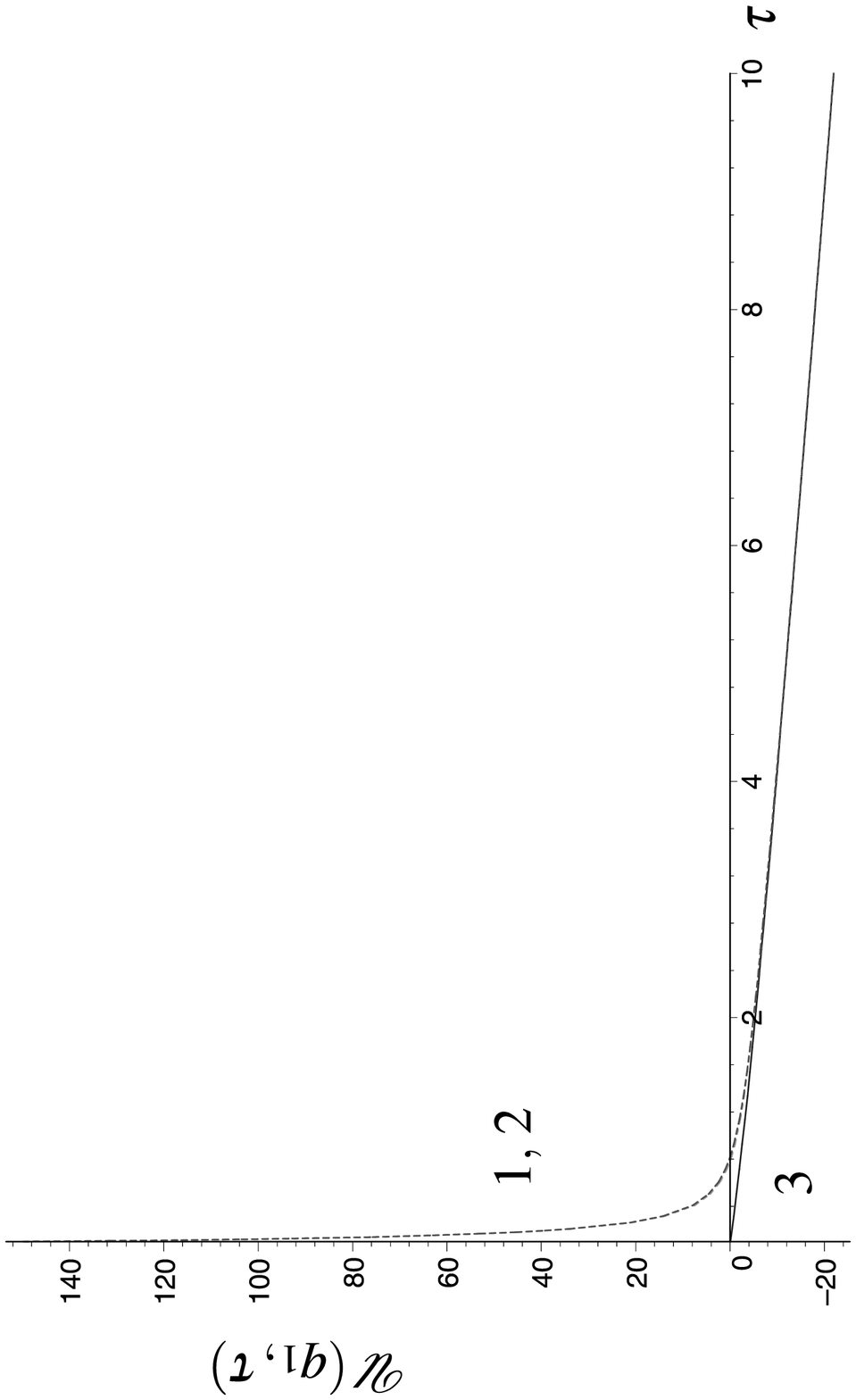}{0.30}{View of the potential $\mathcal U(\tau)$ [Eq.
\eqref{poten}] as a function of $\tau$ corresponding to three different cases.}
{0.45}{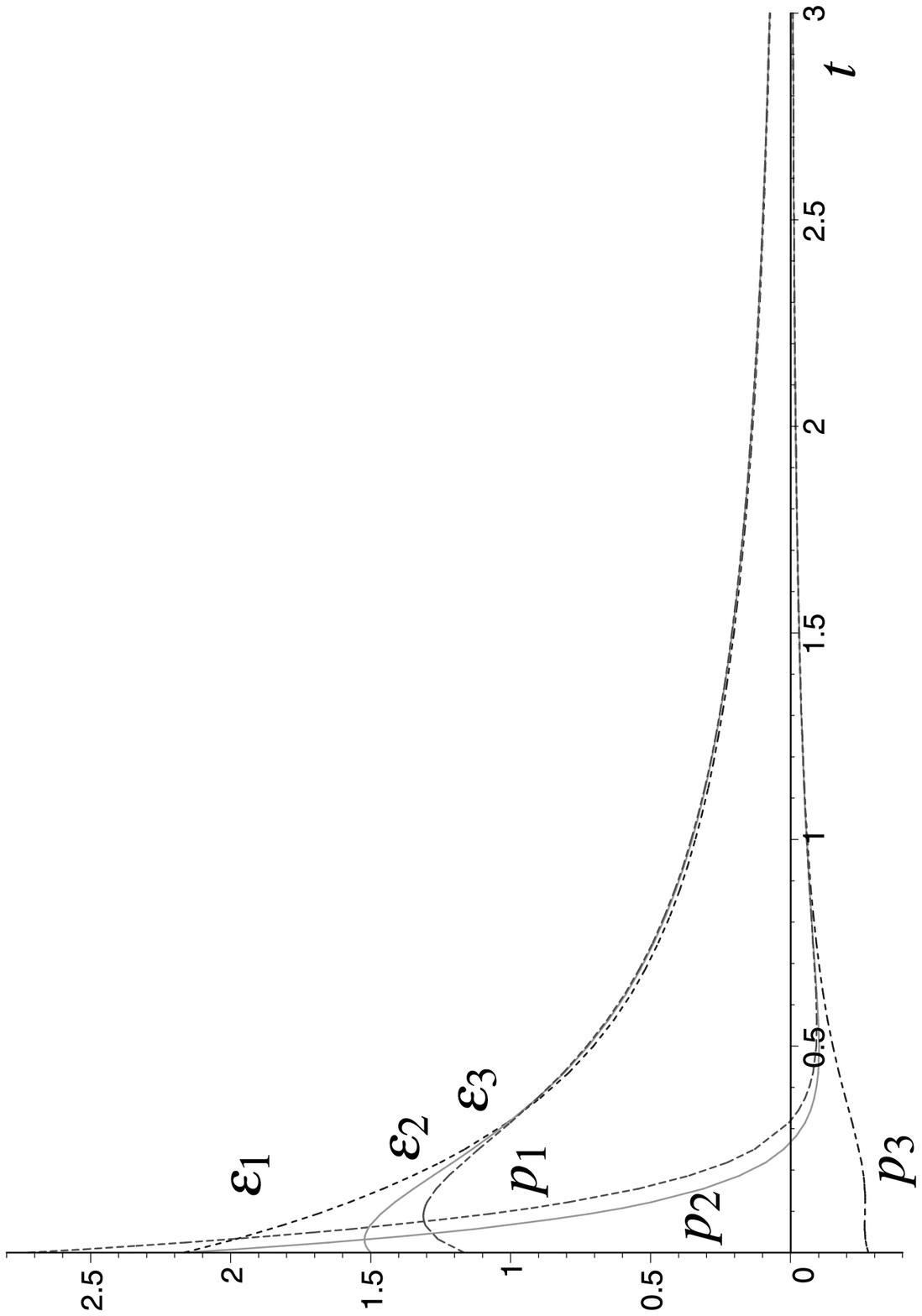}{0.30}{Effective energy density and effective pressure corresponding
to three different cases.}{0.45}

As one sees from Fig. \ref{potenf}, in presence of a self-action of the spinor field,
there occurs an infinitely high barrier as $\tau \to 0$, it means that in the case
considered here $\tau$ cannot be trivial [if treated classically, the Universe cannot
approach to a point unless it stays at an infinitely high energy level].
Thus we see, the nonlinearity of the spinor
field provided by the self-action generates singularity-free evolution of the Universe.
But, as it was shown in \cite{sprd}, this regularity can be achieved only at the expense
of dominant energy condition in Hawking-Penrose theorem. It is also clear that if the
nonlinearity is induced by a scalar field, $\tau$ may be trivial as well, thus giving rise
to space-time singularity. It should be noted that introduction of a positive $\Lambda$
just accelerates the speed of expansion, whereas, a negative $\Lambda$ depending of the
choice of $E$ generates oscillatory or non-periodic mode of evolution. These cases are
thoroughly studied in \cite{sprd,bited}. As it was shown in \cite{bited} the regular solution
obtained my means of a negative $\Lambda$ is case of interaction does not result in broken
dominant energy condition. In Fig. \ref{enpr} we plot the effective energy density and effective
pressure of the matter field. In case of self-action pressure is initially positive,
but with the expansion of the Universe it becomes negative. In case of interaction field the pressure
is always negative. It means, the models with nonlinear spinor field and interacting spinor and scalar
fields can to some extent explain the late time acceleration of the Universe. As one sees, the
corresponding quantities (potential, energy density and pressure) differs only at the initial stage
depending on the type of nonlinearity.

\vskip 1.5 cm

\myfigs{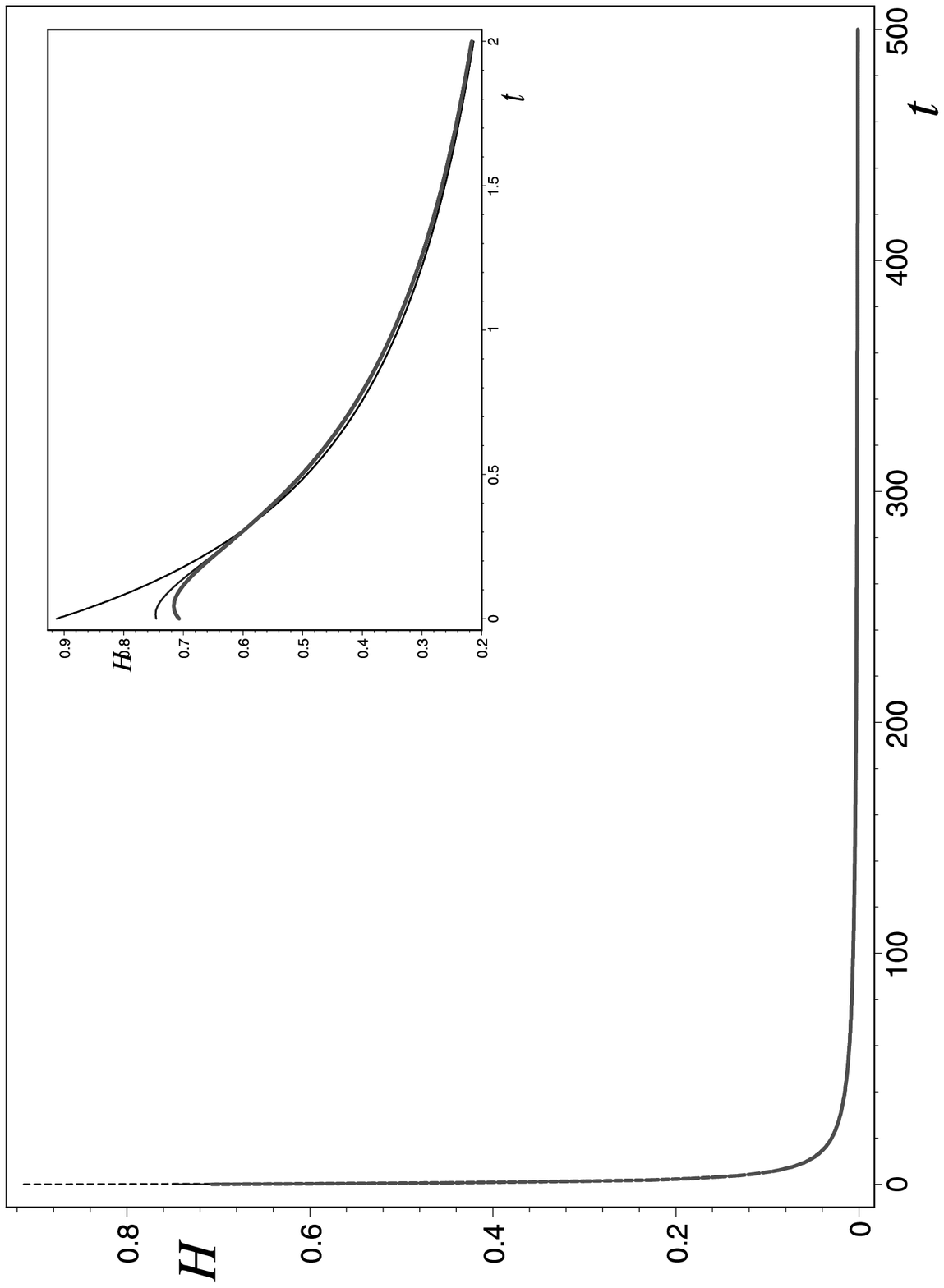}{0.26}{Evolution of the Hubble constant as the Universe expands.}
{0.44}{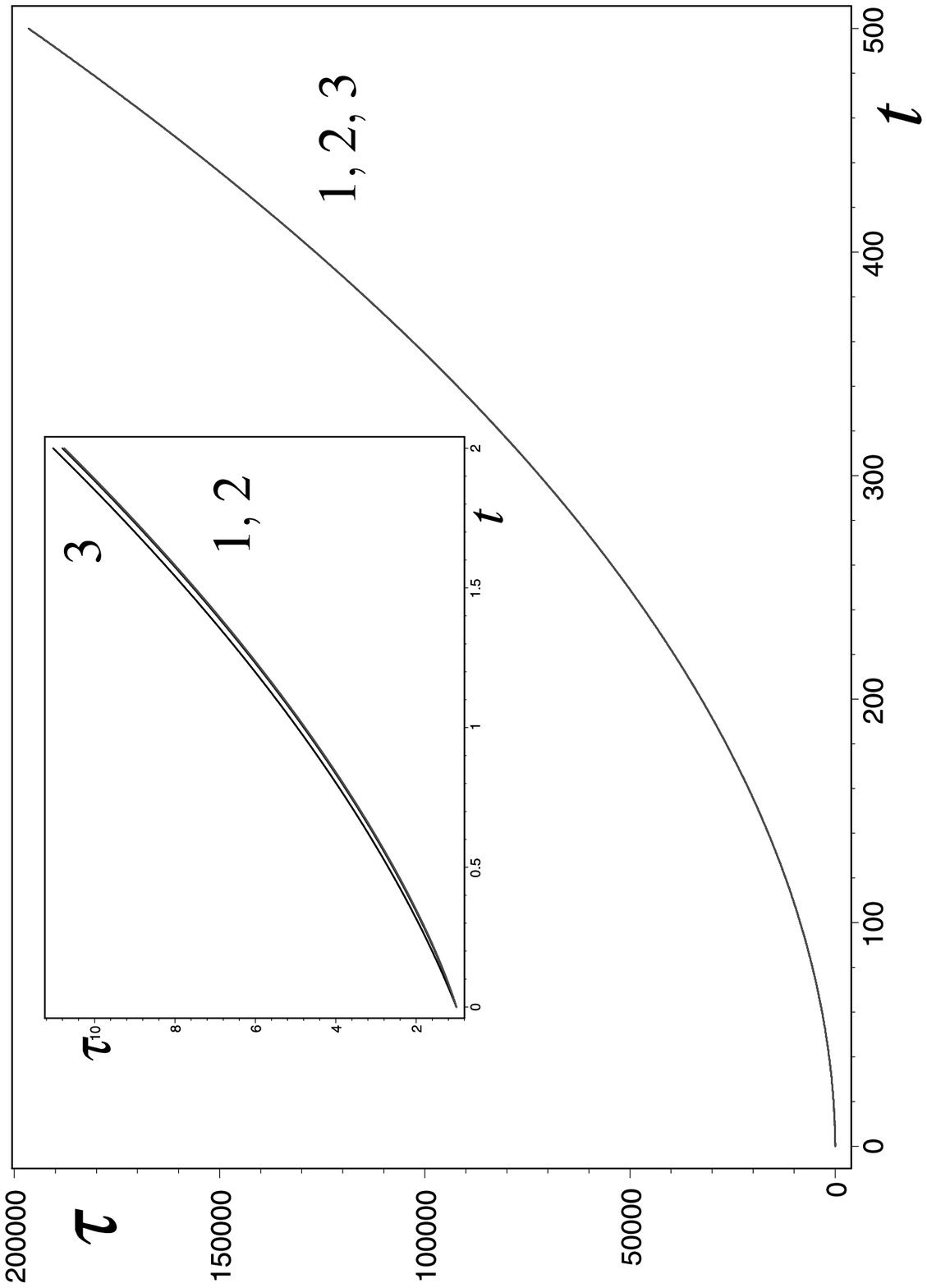}{0.30}{Expansion of the BI Universe with time for three different cases.}{0.44}

In Fig. \ref{H} we graphically justified our assumptions about the role of $\Lambda$ term,
namely, in absence of the cosmological constant, $H \to 0$ as the Universe expands. Finally,
in Fig. \ref{tauf} we illustrate the evolution of the Universe. As one sees, the character of
evolution differs only at the initial stage depending on the choice of nonlinearity.

Finally we would like to emphasize that here we restrict within three cases only. Cases with
nontrivial $\Lambda$ term is not considered, since they were thoroughly studied in previous
papers \cite{lambda,sprd,bited}. Our main aim here was to emphasize the new role of spinor field
to explain the late time acceleration of the Universe.

\section{conclusion}

We considered a system of interaction nonlinear spinor and scalar fields within the scope of a
BI cosmological model filled with perfect fluid. It is  shown that the spinor field nonlinearity
can generate a negative effective pressure, which can be seen as an alternative source for
late time acceleration of the Universe. Here, beside spinor and scalar fields, we consider usual
perfect fluid obeying $p_{pf} = \zeta \ve_{pf}$. We plan to consider a few other fluids in near future
that can provide an initial inflation as well.


\begin{thebibliography}{99}

\bibitem{PRpadma} Padmanabhan, T., Phys. Rep. {\bf 380} (2003) 235.

\bibitem{sahni} Sahni, V., {\it Dark Matter and Dark Energy} arXiv (2004)
astro-ph/0403324.

\bibitem{lambda} Saha, Bijan,
{\it Anisotropic cosmological models with a perfect fluid and a
$\Lambda$ term} (accepted for publication in Astrophysics and
space science) arXiv (2004)
\hnl{gr-qc/0411080}{http://xxx.lanl.gov/abs/gr-qc/0411080}.

\bibitem{caldwell} Cladwell, R.R., Dave, R., and Steinhardt, P.J.,
Phys. Rev. Lett. {\bf 80} (1998) 1582.

\bibitem{starobinsky} Sahni, V. and Starobinsky, A.A., Int. J. Mod.
Phys. D {\bf 9} (2000) 373.

\bibitem{zlatev} Zlatev, I., Wang, L., and Steinhardt, P.J., Phys.
Rev. Lett. {\bf 82} (1999) 896.

\bibitem{pfdenr} Saha, Bijan,
{\it Anisotropic cosmological models with  perfect fluid and dark
energy revisited} arXiv (2005)
\hnl{gr-qc/0501067}{http://xxx.lanl.gov/abs/gr-qc/0501067}.

\bibitem{kamen} Kamenshchik, A.Yu., Moschella, U., and  Pasquier, V.,
Phys. Lett. B. {\bf 511} (2001) 265.

\bibitem{pfden} Saha, Bijan,
{\it Anisotropic cosmological models with perfect fluid and dark
energy} (accepted for publication in Chinese Journal of Physics)
arXiv (2004) \hnl{gr-qc/0412078}{http://xxx.lanl.gov/abs/gr-qc/0412078}.

\bibitem{sjmp} Saha, B. and Shikin, G.N., J. Math. Phys. {\bf 38} (1997)
\hnl{5305}{http://www.jinr.ru/~bijan/my_papers/JMP05305.pdf}.

\bibitem{sgrg} Saha, B. and Shikin, G.N., Gen. Relativ. Gravit. {\bf 29} (1997)
\hnl{1099}{http://www.jinr.ru/~bijan/my_papers/grg97_1099.pdf}.

\bibitem{smpla} Saha, Bijan, Mod. Phys. Lett. A {\bf 16} (2001),
\hnl{1287}{http://www.jinr.ru/~bijan/my_papers/mpla01_1287.pdf}.

\bibitem{sprd} Saha, Bijan, Phys. Rev. D {\bf 64} (2001)
\hnl{123501}{http://www.jinr.ru/~bijan/my_papers/PRD23501.pdf}.

\bibitem{bited} Saha, Bijan and Boyadjiev, T.,
Phys. Rev. D {\bf 69} (2004)
\hnl{124010}{http://www.jinr.ru/~bijan/my_papers/PRD24010.pdf}.

\bibitem{green} Armend$\acute a$riz-Pic$\acute o$n, C. and Greene, P.B.,
Gen. Relativ. Gravit. {\bf 35} (2003) 1637.

\bibitem{kremer1} Ribas, M.O., Devecchi, F.P., and  Kremer, G.M.,
{\it Fermions as sources of accelerated regimes in cosmology} ArXiv
(2005) gr-qc/0511099.

\end{thebibliography}
\end{document}